\documentclass[aps,pra,twocolumn,amsmath,amssymb,nofootinbib,superscriptaddress]{revtex4-1}

\newcommand{\eqnref}[1]{Eq.~(\ref{#1})}

\newcommand{\figref}[1]{Fig.~\ref{#1}}

\newcommand{\bra}[1]{\langle#1|}
\newcommand{\ket}[1]{|#1\rangle}

\newcommand{\abs}[1]{\left|#1\right|}
\newcommand{\e}{\mathrm{e}}
\newcommand{\ii}{\mathrm{i}}

\newcommand{\mean}[1]{\left\langle #1 \right\rangle } % mean
\usepackage{graphicx}
\usepackage[colorlinks]{hyperref}

\begin{document}

\title{Conclusive Precision Bounds for SU(1,1) Interferometers}

\author{Chenglong You}
\email[]{cyou2@lsu.edu}
\affiliation{Hearne Institute for Theoretical Physics and Department of Physics \& Astronomy, Louisiana State University, Baton Rouge, LA 70803, United States}
\affiliation{National Institute of Information and Communications Technology, Koganei, Tokyo 184-8795, Japan}

 \author{Sushovit Adhikari}
 \affiliation{Hearne Institute for Theoretical Physics and Department of Physics \& Astronomy, Louisiana State University, Baton Rouge, LA 70803, United States}
 \affiliation{National Institute of Information and Communications Technology, Koganei, Tokyo 184-8795, Japan}

 \author{Xiaoping Ma}
 \affiliation{Hearne Institute for Theoretical Physics and Department of Physics \& Astronomy, Louisiana State University, Baton Rouge, LA 70803, United States}
 \affiliation{Department of Physics, Ocean University of China, Qingdao, 266100, China}

 \author{Masahide Sasaki}
 \affiliation{National Institute of Information and Communications Technology, Koganei, Tokyo 184-8795, Japan}
 
 \author{Masahiro Takeoka}
 \affiliation{National Institute of Information and Communications Technology, Koganei, Tokyo 184-8795, Japan}

\author{Jonathan P. Dowling}
 \affiliation{Hearne Institute for Theoretical Physics and Department of Physics \& Astronomy, Louisiana State University, Baton Rouge, LA 70803, United States}
 \affiliation{National Institute of Information and Communications Technology, Koganei, Tokyo 184-8795, Japan}
 \affiliation{NYU-ECNU Institute of Physics at NYU Shanghai,
 3663 Zhongshan Road North, Shanghai, 200062, China}
 \affiliation{CAS-Alibaba Quantum Computing Laboratory, CAS Center for Excellence in Quantum Information and Quantum Physics, University of Science and Technology of China, Shanghai 201315, China}

\date{\today}

\begin{abstract}
In this paper, we revisit the quantum Fisher information (QFI) calculation in SU(1,1) interferometer considering different phase configurations. When one of the input modes is a vacuum state, we show by using phase averaging, different phase configurations give same QFI. In addition, by casting the phase estimation as a two-parameter estimation problem, we show that the calculation of the quantum Fisher information matrix (QFIM) is necessary in general. Particularly, within this setup, the phase averaging method is equivalent to a two parameter estimation problem. We also calculate the phase sensitivity for different input states using QFIM approach. 
\end{abstract}

\maketitle

\section{Introduction}
Metrology, like other fields in physics, has been re-investigated with respect to the laws of quantum mechanics. The fundamental setup of a metrology device is a Mach-Zehnder interferometer (MZI). MZIs are widely used in the study of phase estimation. By utilizing only classical resources, the precision of estimation or sensitivity of this device is bounded by $\Delta^2 \phi \ge 1/\bar{n}$, where $\bar{n}$ is the average photon number inside the interferometer. This bound is referred to as the shotnoise limit (SNL) \cite{caves81}. 
However, this is not the ultimate limit. If one deploys quantum resources, then one can beat the SNL and reach the Heisenberg limit (HL), $\Delta^2 \phi \ge 1/\bar{n}^2$ \cite{caves81}. The estimation precision of the interferometer can be given by the error propagation formula for a given measurement scheme \cite{kay1993}. However, this approach cannot determine the optimal sensitivity without optimizing over all possible measurement schemes. To circumvent this problem, Braunstein and Caves introduced quantum Fisher information (QFI), which depends only on the input state and not on a particular measurement scheme \cite{QFI_Caves}. 
The ultimate precision bounds of the phase sensitivity is then given by the quantum Cra\'mer-Rao (QCRB) bound
\cite{QFI_Caves,Helstrom1976,Caves1993}, $\Delta^2 \phi \ge 1/F_Q$, where $F_Q$ is the QFI. 
%This sets the lower bound on the precision of estimation \cite{Helstrom1976,Caves1993}.

For MZI, where all elements in the setup are passive, the SNL can be beaten by use of exotic quantum states: squeezed states \cite{caves81}, N00N states \cite{Boto2000}, twin Fock states \cite{Dowling2008} and two-mode squeezed states \cite{Anisimov2010}. The other way of beating SNL is to use active elements in an interferometer. One of such interferometer is called a SU(1,1) interferometer, introduced by Yurke in 1986 \cite{Yurke1986}, where the beam splitters in the MZI are replaced by an active element, such as an optical parametric amplifier (OPA) or a four-wave mixers, which are mathematically characterized by the group SU(1,1). 
These interferometers can achieve Heisenberg sensitivity even if inputs are both vacua. Since first proposed, the phase sensitivity of the SU(1,1) interferometer has been extensively studied both in theory and experiment. Plick \textit{et al.}~\cite{Plick2010} showed that coherent state inputs with intensity measurement could achieve higher sensitivity which was experimentally demonstrated by Ou \cite{Ou12}. 
Li \textit{et al.}~\cite{Li2014,Li2016} calculated the phase sensitivity of the SU(1,1) interferometer with coherent and squeezed vacuum states as inputs and  homodyne and parity measurement as detection when the unknown phase shift is applied in one of the arms. They gave the QFI-based analysis as well. 
Gong \textit{et al.}~\cite{Gong2016} also did the QFI analysis for coherent and squeezed states where the unknown phase shifts are applied in both arms. More recently, there has been an interest in other variant termed as pumped-up and truncated SU(1,1) \cite{Paullet1,Paullet2,Szigeti2017,Zhang2018}. 

One issue in the SU(1,1) interferometer is that the QFI gives different precision limits with different configurations of the unknown phases, even for the same physical setup. 
For example, as mentioned recently by Gong et al. \cite{Gong2016}, three different phase configurations would yield three different QFIs with the same setup for most of the Gaussian state inputs. 
This issue was also observed in the MZI setting, and
Jarzyna and Demkowicz-Dobrza\'{n}ski \cite{Rafal2012} pointed out that without proper consideration of external phase reference, the physically same setup  lead to different QFIs in MZI. 
That is, naive calculation of QFI sometimes misleadingly overestimates the precision limit in the sense that, to achieve it, one requires {\it hidden} uncounted resource in measurement. 
They also proposed a technique to rule out these hidden resources by averaging the phase of the input states. 
Related to this, a rigorous justification of the fundamental precision limit of the MZI when one of the inputs is a vacuum 
was recently discussed \cite{Takeoka}. 

In this paper, we revisit the QFI analysis of the SU(1,1) interferometer with various input states and unknown-phase models. 
To avoid overestimation of the precision limit, we apply the two approaches, the phase averaging of the input 
and the quantum Fisher information matrix (QFIM). 
First, we consider the SU(1,1) interferometer where one of the inputs is restricted to be a vacuum. 
We give a tight QCRB for the one- and two-unknown-phase models. For the former, we apply the phase averaging technique to rule out the use of external resources at the measurement, and for the latter, we use the multi-parameter estimation approach, and the bound we derive is valid even allowing external resources. 
We also consider two non-vacuum inputs and give tighter bounds than previously reported. 

\section{Model and Previous Work}

%%%%%%%%%%Schematic%%%%%%%%%%%%
\begin{figure}[t]
\centering
\includegraphics[width=0.4\textwidth]{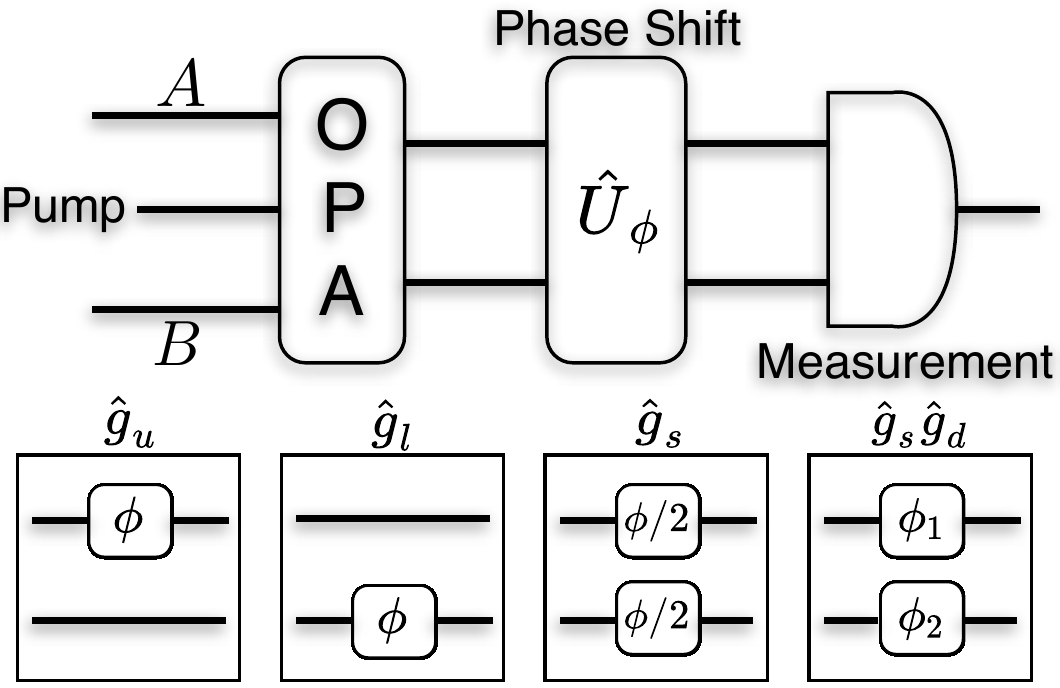}
\caption{Schematic of a SU(1,1) interferometer with different phase shift models: phase shift in upper arm only ($\hat{g}_u$ model), phase shift in lower arm only ($\hat{g}_l$ model), phase shift equally splitted in both arms ($\hat{g}_s$ model), unknown phase shifts in both arms ($\hat{g}_s-\hat{g}_d$ model).}
\label{schematic}
\end{figure}
%%%%%%%%%%Schematic%%%%%%%%%%%%

\subsection{Model}
A schematic of the SU(1,1) interferometer is shown in \figref{schematic}. Two input modes interact via optical parametric amplifier (OPA) with gain parameter $g$, and then go through a phase shift on one or both of the arms. After the phase shifts, the measurement is performed. 
Note that the measurement often consists of a second OPA with pumping of $\pi$ phase difference than the first one followed by detectors. 
%Here, we restrict one of the input states to be a vacuum state $\ket{0}$, whereas the other input can be an arbitrary pure state $\ket{\chi}$. 

The relation between the output and input modes of the OPA is given by:
\begin{align}
\label{eqn:inoutrelation}
{\hat{a}_1} &= \cosh(g) \, {\hat{a}_0} + {\e^{\ii{\theta }}}\sinh(g) \, \hat{b}_0^\dag, \\ \nonumber
\hat{b}_1 &= \cosh(g) \, \hat b_0 + {\e^{\ii{\theta }}}\sinh(g) \, \hat a_0^\dag,
\end{align}
where $\hat{a}_i$ ($\hat{b}_i$) and $\hat{a}_i^\dagger$ ($\hat{b}_i^\dagger$) are the annihilation and creation operators in mode A (B), respectively, and the subscripts 0 and 1 represent the input and the output of the OPA, respectively (see \figref{schematic}).
$g$ and $\theta$ are the parametric gain and phase of the OPA, respectively.

The unknown phase shifts to be estimated are modeled in different ways. 
The choice of the models depend on what type of application scenario one has in mind \cite{Takeoka}. 
When the unknown phase shift $\phi$ occurs only in the upper arm, it is modeled by unitary operation $\hat{U}_{\phi}^{u} = \e^{\ii \hat{g}_{u} \phi}$ with generator $\hat{g}_{u} = \hat { a }_{1}^{\dagger}\hat { a }_{1}$. 
If the phase shift is in only in the lower arm, we have $\hat{U}_{\phi}^l = \e^{\ii \hat{g}_{l} \phi}$ with generator $\hat{g}_{l} = \hat { b }_{1}^{\dagger}\hat { b }_{1}$. 
Sometimes, the unknown phase shift is equally split into two arms, then $\hat{U}_{\phi}^s = \e^{\ii \hat{g}_{s} \phi}$ where $\hat{g}_{s} = (\hat { a }_{1}^{\dagger}\hat { a }_{1} + \hat { b }_{1}^{\dagger}\hat { b }_{1})/2$. 
In some applications, different unknown phase shifts occur in the each arms. The unitary operator is then given by $\hat{U}_{\phi} = \e^{\ii \hat{g}_{1}\phi_{1}}\e^{\ii \hat{g}_{2}\phi_{2}}=\e^{\ii \hat{g}_{s}\phi_{s}}\e^{\ii \hat{g}_{d}\phi_{d}}$, 
where $\hat{g}_{d} = (\hat { a }_{1}^{\dagger}\hat { a }_{1} - \hat { b }_{1}^{\dagger}\hat { b }_{1})/2$ and $\phi_1$ and $\phi_2$ are the unknown phases in the two arms. These phases are also described by the phase sum $\phi_{s}=\phi_{1} + \phi_{2}$, and the phase difference $\phi_{d}=\phi_{1}-\phi_{2}$. 
Note that the first three cases are basically a single-parameter estimation problem and the last one is a two-parameter estimation problem.

\subsection{Review of the previous QFI approach}

Here we review the previous QFI approach to determine the sensitivity bound for the single-parameter 
SU(1,1) interferometric sensing. 
When the state before the measurement is a pure state, $|\psi_{\phi} \rangle_{AB} = e^{-\ii \hat{g} \phi} |\psi \rangle_{AB}$, the QFI is given by \cite{Caves1993}
 \begin{equation}
 \label{eqn:QFIcalc}
 F_Q = 4(\bra{\psi}\hat{g}^2\ket{\psi}-\bra{\psi}\hat{g}\ket{\psi}^2).
 \end{equation}
% Moreover, if we restrict the input states to be Gaussian states, then all the operations in SU(1,1) interferometer could be described by the transformation of the mean and covariance of the input states. Specifically, the QFI is given by \cite{Monras}
% \begin{equation}
% \label{eqn:QFIcalc}
% F = \frac{1}{4} \Tr[\dot \Gamma {\Gamma ^{ - 1}}\dot \Gamma {\Gamma ^{ - 1}}] + 2{\dot d^\top}{\Gamma ^{ - 1}}\dot d,
% \end{equation}
% where $\Gamma$ and $d$ is the covariance and mean after the transformation, and $\dot O\equiv \partial O/\partial \phi$. 
By using \eqnref{eqn:inoutrelation} and \eqnref{eqn:QFIcalc}, when the input states are both vacuum states, all different phase configurations yield the same QFI: 
\begin{equation}
\label{eqn:vacuuminput}
 F_Q=n_{\kappa}(n_{\kappa}+2),
\end{equation}
 where $n_{\kappa}=2\sinh^2(g)$ is the average photon number of the output state. 
That is, the precision bound is independent of what model one chooses. 
%This result suggests that regardless of what the phase configuration is, the precision bound of the SU(1,1) interoferemeter is always the same. 
However, the situation starts to change when one of the input mode is not a vacuum state. 
 
For example, in Ref.~\cite{Gong2016}, the authors calculated the QFI from \eqnref{eqn:QFIcalc}, where one input is a coherent state and the other is a coherent or squeezed state. To simplify, taking one input to be coherent state $\ket{\beta}$ with $n_\beta=\abs{\beta}^2$ and the other to be a vacuum, the QFIs in Ref.~\cite{Gong2016}, for generators $\hat{g}_u$, $\hat{g}_l$, and $\hat{g}_s$ are reduced to,
\begin{eqnarray}
\label{eqn:f_q(u)}
F_Q(\hat{g}_u) & = & n_\beta \cosh{4g} + \sinh^2{2g} + n_\beta(1-2\cosh{2g}),
\\
\label{eqn:f_q(l)}
F_Q(\hat{g}_l) & = & n_\beta \cosh{4g} + \sinh^2{2g} + n_\beta(1+2\cosh{2g}),
\\
\label{eqn:f_q(s)}
F_Q(\hat{g}_s) & = & n_\beta \cosh{4g} + \sinh^2{2g}, 
\end{eqnarray}
which suggests that a choice of the phase shift model changes the QCRB of the phase estimation even using the same physical setup. 
Then one may ask a question: do they reflect tight precision limits? 
In other words, is there any possibility of overestimating the bound by use of {\it hidden} resources at the measurement?
In the next section, we show that after ruling out any external resources, the QCRBs are unified and tighter than any of the above bounds.

\section{Single-phase estimation with a vacuum in one input and an arbitrary state in the other}
In this section, we consider a slightly more general situation than that in the previous section. 
We consider the single-phase estimation of the SU(1,1) interferometer where 
one input (mode A) is an arbitrary state $\hat{\rho}_\chi$ and the other (mode B) is a vacuum $|0\rangle\langle0|$. 

As pointed out in Refs. \cite{Rafal2012, Takeoka}, the QFI-only approach sometime falsely suggest a quantum advantage, as the optimal measurement saturating the bound may include uncounted resources, such as an external strong local oscillator. 
The possibility of this overestimation is circumvented by eliminating the common reference frame between the input states and the measurement. 
Let us expand $\hat{\rho}_\chi$ in the photon-number basis: 
%%%%%%%%%%%%%%%%%%%%%%%%%%%
\begin{equation}
\label{eq:input_state}
\hat{\rho}_\chi = \sum_{n,m=0}^\infty c_{nm} |n \rangle\langle m|,
\end{equation}
%%%%%%%%%%%%%%%%%%%%%%%%%%%
where $|n\rangle$ is the $n$-photon number state. 
The reference frame between the inputs and the measurement are removed by phase-averaging the input state as \cite{Rafal2012}, 
%%%%%%%%%%%%%%%%%%%%%%%%%%%
\begin{align}
\label{eq:phase_averaged_input}
\Psi_{\textrm{avg}} &= 
\int\frac{d\varphi}{2\pi} \hat{V}^A_\varphi \hat{V}^B_\varphi 
\left( \hat{\rho}_\chi^A \otimes |0 \rangle\langle 0|^B \right) 
\hat{V}^{A \, \dagger}_\varphi \hat{V}^{B \, \dagger}_\varphi 
\nonumber\\ &= 
\sum_{n,m=0}^{\infty}\int\frac{d\varphi}{2\pi}
e^{i\varphi (n-m)} c_{nm} |n\rangle \langle m|^A 
\otimes |0\rangle \langle 0|^B 
\nonumber\\ &= 
\sum_{n=0}^{\infty} p_n
|n\rangle \langle n|^A \otimes |0\rangle \langle 0|^B ,
\end{align}
%%%%%%%%%%%%%%%%%%%%%%%%%%%
where $\hat{V}^A_\varphi = e^{i\varphi \hat{a}^\dagger \hat{a}}$, 
$\hat{V}^B_\varphi = e^{i\varphi \hat{b}^\dagger \hat{b}}$, and 
$p_n = c_{nn}$ is a real positive number satisfying 
$\sum_n p_n =1$ \cite{Takeoka, AgarwalBook}.

The state after the first OPA is given by
%%%%%%%%%%%%%%%%%%%%%%%%%%%
\begin{align}
\label{eq:phase_averaged_output}
\Psi_{\textrm{avg}}^{\textrm{OPA}} &=
\hat{T}^{AB}_{\textrm{OPA}} \Psi_{\textrm{avg}}
\hat{T}_{\textrm{OPA}}^{\dagger AB}
\nonumber\\ &= 
\sum_{n=0}^{\infty} p_n
\left( \hat{T}^{AB}_{\textrm{OPA}} |n\rangle \langle n|^A \otimes |0\rangle \langle 0|^B
\hat{T}_{\textrm{OPA}}^{\dagger AB} \right )
\nonumber\\ &= 
\sum_{n=0}^\infty 
p_n |\psi_n\rangle\langle\psi_n|_{AB} ,
\end{align}
%%%%%%%%%%%%%%%%%%%%%%%%%%%
where 
%%%%%%%%%%%%%%%%%%%%%%%%%%%
\begin{equation}
|\psi_n\rangle _{AB}=\sum\limits_{k = 0}^\infty {{c_{n,k}}{{\left| {k + n} \right\rangle }_A} \otimes {{\left| k \right\rangle }_B}}, 
\end{equation}
%%%%%%%%%%%%%%%%%%%%%%%%%%%
which follows from the fact that the photon number difference between two arms is conserved \cite{Yurke1986}.
Note that $c_{n,k}$ is a function of the OPA phase $\theta$. 
This means that we only average the input phases but not the OPA phase $\theta$. 
That is, we allow one to use the reference frame $\theta$ and corresponding extra resources at the measurement step as it is in the original proposal of the SU(1,1) interferometric sensing \cite{Yurke1986}.

By using the convexity of the QFI \cite{fujiwara2008,demkowicz2015} and noticing that 
$|\psi_n\rangle$ and $|\psi_{n'}\rangle$ 
are orthogonal for $n \ne n'$, we have 
%%%%%%%%%%%%%%%%%%%%%%%%%%%
\begin{equation}
F_{Q}\left(\Psi_{\textrm{avg}}\right)
= \sum_{n=0}^{\infty} p_n 
F_{Q}\left(|\psi_n\rangle \right) .
\end{equation}
%%%%%%%%%%%%%%%%%%%%%%%%%%%

%Different from the MZI case, \eqnref{eq:phase_averaged_output} consists of infinite sum, thus we use Heisenberg picture to calculate the QFI for the phase-averaged input state. 
For the phase shift in upper arm only model $\hat{g}_u$, the QFI of $|\psi_n\rangle$ is given by
\begin{equation}
 F^u_{Q}\left(|\psi_n\rangle \right) = 4(\mean{\hat{g}_u^2}-\mean{\hat{g}_u}^2),
 \end{equation}
where 
\begin{equation}
\mean{\hat{g}_u} = n\cosh^2(g) + \sinh^2(g),
\end{equation}
and 
\begin{align}
\mean{\hat{g}_u^2}&=n^2\cosh^4(g)+(n+1)\cosh^2(g)\sinh^2(g)\nonumber
\\ &+2n\cosh^2(g)\sinh^2(g)+\sinh^4(g).
\end{align}
Thus,
\begin{equation}
F^u_{Q}\left(|\psi_n\rangle \right) = 4(n+1)\sinh^2(g)\cosh^2(g) = (n+1)n_{\kappa}(n_{\kappa}+2).
\end{equation}
Consequently, the QFI of the phase-averaged input state is given by
\begin{align}
\label{eq:trueQFI}
 F^u_{Q}\left(\Psi_{\textrm{avg}}\right)
&= \sum_{n=0}^{\infty} p_n (n+1)n_{\kappa}(n_{\kappa}+2)\nonumber
\\&=(\bar{n}_\chi+1)n_{\kappa}(n_{\kappa}+2), 
\end{align}
where $\bar{n}_\chi = \sum_n n p_n$ is the average photon number of $\hat{\rho}_\chi$. 

Similarly, for the phase shift in lower arm only model $\hat{g}_l$, 
the phase-averaged input state yields, 
 \begin{equation}
\label{f^l}
 F^l_{Q}\left(\Psi_{\textrm{avg}}\right) =(\bar{n}_\chi+1)n_{\kappa}(n_{\kappa}+2).
 \end{equation}
Finally, for the equally-split phase shift model $\hat{g}_s$ we get, 
 \begin{equation}
\label{f^s}
 F^s_{Q}\left(\Psi_{\textrm{avg}}\right) =(\bar{n}_\chi+1)n_{\kappa}(n_{\kappa}+2).
 \end{equation}
The results in Eqs.~(\ref{eq:trueQFI}--\ref{f^s}) show that, once we rule out the use of external references (resources) at the measurement, all these models give the same QFI. 
Note that by taking $\bar{n}_\chi \to 0$, i.e. setting the two inputs both vacuum, we recover Eq.~(\ref{eqn:vacuuminput}). 

When the input state in mode A is a coherent state $\ket{\alpha}$, we have
\begin{equation}
\label{eq:CohTrueQFI}
 F_Q^\alpha=(n_\alpha +1)n_{\kappa}(n_{\kappa}+2).
\end{equation}
This is tighter than any QFIs without phase-averaging mentioned in the previous section [Eqs.~(\ref{eqn:f_q(l)}--\ref{eqn:f_q(s)})]. 
That is, to achieve the QFIs in Eqs.~(\ref{eqn:f_q(l)}--\ref{eqn:f_q(s)}), external (but uncounted) resources at the measurement are required. 
Note that our bound in \eqnref{eq:CohTrueQFI} is tight in the sense that it is saturated by the parity detection \cite{Li2016}.

From \eqnref{eq:trueQFI} we can see that, when one of the input is a vacuum, the QFI is proportional to $\bar{n}_\chi$ but does not depend on the structure of $\hat{\rho}_\chi$. 
This means that if we fix the OPA gain $g$, then the best strategy is to use a state with higher average photon number. 
In other words, no nonclassicality of the input state can boost the sensitivity, when the other input is a vacuum. 
This situation has some similarity with the MZI case \cite{Takeoka}, where if one of the inputs is vacuum, one cannot beat the shotnoise limit by any nonclassical input from the other port. 
Also, if we restrict the total amount of resources used in both the input state and the OPA, then the best strategy is to concentrate all the power resource into the OPA.
%From \eqnref{eq:trueQFI} we can see the QFI of this particular input states is linear proportional to the average photon number $\bar{n}_{\chi}$ of the arbitrary state $|\chi\rangle$, and quadratic proportional to the average photon number $n_{\kappa}$ of the two mode squeezer (equivalently, OPA). This suggests that if we fix the squeezing strength $g$ of the OPA, then the best strategy is to put a state with higher average photon number. In other word, exotic quantum states won't help, which is similar to the shotnoise limit in the MZI case. If we consider both input state and OPA as resources, then the best strategy is to increase the squeezing strength of the OPA. 

\section{Two-phase estimation with a vacuum in one input and an arbitrary state in the other}

In the last section, we consider single-phase estimation and use the phase-averaging method to calculate the QFI of the SU(1,1) interferometer, where the input state is given by an arbitrary state and a vacuum state. 
For the $\hat{g}_u$ and $\hat{g}_l$ models, all the phase shift is located in only one of the arms.
In other words, we assume that we know there is no phase shift in the other arm. 
%Therefore $\hat{g}_1$ and $\hat{g}_2$ model is a single parameter estimation problem. 
The $\hat{g}_s$ only model is also a single-parameter estimation since we assume the common unknown phase shift occurs in both arms. That is, we (implicitly) assume that the phase difference $\phi_d = \phi/2 - \phi/2 =0$ is known {\it a priori}. 
However, in many applications, different phase shift occurs in each arm, or equivalently, both the phase sum ($\phi_s$) and phase difference ($\phi_d$) are unknown. 
Then one needs to consider the two-parameter estimation problem, even if only one of them is of interest. 
%one would argue that there are phase shifts in both arm, therefore it will become a two-parameter estimation problem. 
%This is not true since the one still has information about the relationship between phase shifter in two arms. In fact, in all three models we mentioned above, we implicitly assumed that $\phi_d$ is priori known. Therefore, the correct approach to the SU(1,1) phase estimation problem is to treat the scheme as a two-parameter estimation problem and calculate the precision limit from the multiparameter QCRB. 

In this section, we consider the estimation of $\phi_s$ and $\phi_d$ with the SU(1,1) interferometer where again one input is an arbitrary state and the other is a vacuum. 
Without loss of generality, we can restrict the former to be a pure state $|\chi\rangle$ 
(this is justified by the convexity of the quantum Fisher information quantities, for example, see Ref. \cite{Takeoka}).

The QCRB for multi-parameter estimation is calculated through the quantum Fisher information matrix (QFIM) \cite{Helstrom1976,you2017}. 
For the estimation of $\phi_d$ and $\phi_s$, its QFIM is given by a two-by-two matrix: 
\begin{equation}
\mathcal{F}_Q=\left[ {\begin{array}{*{20}{c}}
{{F_{dd}}}&{{F_{sd}}}\\
{{F_{ds}}}&{{F_{ss}}}
\end{array}} \right] ,
\end{equation}
where $F_{ij} = 4 \left( \langle \hat{g}_i \hat{g}_j \rangle - \langle \hat{g}_i \rangle \langle \hat{g}_j \rangle \right)$ and the subscripts $s$ and $d$ denote $\phi_s$ and $\phi_d$, respectively. 

In the MZI case, we are usually interested in the estimation of only $\phi_d$. 
Then its QCRB derived from the QFIM is given by \cite{Takeoka,Rafal2012} 
\begin{equation}
\label{eqn:MZI_QFIM}
 \Delta^2 \phi_d\ge \frac{F_{ss}}{F_{dd} F_{ss}-F_{ds} F_{sd}}.
\end{equation}
Only when the beam splitter of the MZI is 50/50, do the off-diagonal elements vanishes as $F_{sd}=F_{ds}=0$, and then Eq. (\ref{eqn:MZI_QFIM}) simplifies to $\Delta^2 \phi_d\ge F_{dd}^{-1}$. 
That is, one does not need to care if $\phi_s$ is known a priori to find the ultimate precision limit of estimating $\phi_d$. 
In general, however, one needs to take into account the full QFIM, i.e. the precision limit of estimating $\phi_d$ depends on the presence or absence of the information of $\phi_s$. 

The same situation is applied to the SU(1,1) interferometer. 
%Here, we calculate the phase sensitivity of the SU(1,1) interferometer by casting it as a two parameter estimation problem. 
Since the only interesting quantity to measure in the SU(1,1) interferometer is the phase sum $\phi_{s}$, 
we are interested in the QCRB of $\phi_s$: 
\begin{equation}
\label{eqn:SU11_QFIM}
 \Delta^2 \phi_s\ge \frac{F_{dd}}{F_{dd} F_{ss}-F_{ds} F_{sd}}. 
\end{equation}
Applying our input states $\ket{\chi} \otimes \ket{0}$ into the QFIM elements for the SU(1,1) interferometer, we have 
\begin{align}
F_{dd} &= V_{\chi},\\ 
F_{ds} &= F_{sd} = V_{\chi}\cosh(2g), \\ 
F_{ss} &= V_{\chi}\cosh^2(2g) +(1+\bar n_\chi) \sinh^2(2g), 
\end{align}
where $\bar n_\chi$ is the average photon number of state $\ket{\chi}$ and $V_\chi = \langle\chi| \hat{n}^2 |\chi\rangle - \langle\chi| \hat{n} |\chi\rangle^2$ is the photon number variance of $|\chi\rangle$. 
Plugging them into \eqnref{eqn:SU11_QFIM}, we get 
\begin{equation}
\label{eqn:onemode_SU11_QFIM}
 \Delta^2 \phi_s\ge \frac{1}{(\bar n_\chi+1)n_{\kappa}(n_{\kappa}+2)}.
\end{equation}
The above calculation reveals a critical role of the non-diagonal terms $F_{ds}$ and $F_{sd}$. 
In fact, they contribute to cancel out $V_\chi$ from Eq.~(\ref{eqn:onemode_SU11_QFIM}). 
As a consequence, the expression of Eq.~(\ref{eqn:onemode_SU11_QFIM}) coincides with the one 
for the single-parameter estimation with the phase averaging [Eqs.~(\ref{eq:trueQFI}--\ref{f^s})]. 
Thus any nonclassicality of $|\chi\rangle$ does not help to boost the sensitivity. 
Also it should be noted that if one ignore the non-diagonal terms, i.e. implicitly assuming that $\phi_d$ is known a priori, 
one could get higher QFI than Eq.~(\ref{eqn:onemode_SU11_QFIM}), which misleadingly overestimates the precision limit. 

%In the case of $\ket{\chi}=\ket{\alpha}$, we get 
%\begin{equation}
% \Delta^2 \phi_s\ge\frac{1}{(n_\alpha +1)n_{\kappa}(n_{\kappa}+2)},
%\end{equation}
%which also agrees with the QCRB given by phase-averaging method.

Finally, we discuss why the QCRB of the single-parameter estimation with the phase-averaging and the two-parameter estimation coincide when one of the inputs is vacuum. 
Let $\rho=\hat{\rho}_{\rm in}^A \otimes |0 \rangle\langle 0|^B$ and consider its phase averaging.
Then we observe, 
%We would like to highlight the equivalence between the phase averaging method and QFIM method, particularly when $\rho=\hat{\rho}_{\rm in}^A \otimes |0 \rangle\langle 0|^B$. Notice that in \eqnref{eq:phase_averaged_output} that we can commute the OPA operator inside the integration, i.e,
\begin{align}
\Psi_{\textrm{avg}}^{\textrm{OPA}}
&=\int\frac{d\theta}{2\pi} \hat{T}^{AB}_{\textrm{OPA}} \hat{V}^A_\theta \hat{V}^B_\theta 
\hat{\rho}
\hat{V}^{A \, \dagger}_\theta \hat{V}^{B \, \dagger}_\theta \hat{T}_{\textrm{OPA}}^{\dagger AB} \nonumber\\ 
&=\int\frac{d\theta}{2\pi} \hat{T}^{AB}_{\textrm{OPA}} \hat{V}^A_\theta \hat{V}^B_{-\theta} 
\hat{\rho}
\hat{V}^{A \, \dagger}_\theta \hat{V}^{B \, \dagger}_{-\theta} \hat{T}_{\textrm{OPA}}^{\dagger AB} \nonumber\\ 
&=\int\frac{d\theta}{2\pi} \hat{V}^A_\theta \hat{V}^B_{-\theta} \hat{T}^{AB}_{\textrm{OPA}} 
\hat{\rho} \hat{T}_{\textrm{OPA}}^{\dagger AB} \hat{V}^{A \, \dagger}_\theta \hat{V}^{B \, \dagger}_{-\theta}, 
\end{align}
where the second equality follows from the fact that the phase shift does not change the vacuum state and 
the third equality holds since the two-mode squeezing (OPA) operation commutes with the phase shift $\hat{V}^A_\theta \hat{V}^B_{-\theta}$. 
This shows that the phase averaging is effectively equivalent to 
adding another unknown phase $\theta$ in upper arm (mode A) and unknown phase $-\theta$ in lower arm (mode B).
That is, the interferometer's phase difference is set to be unknown and thus the problem is equivalent to the estimation of two unknown parameters, $\phi_s$ and $\phi_d = 2\theta$ \cite{Marcin2016}. 
This results in the same precision bound for these two different problems. 
Note that the above only holds when one of the inputs is a vacuum but may not hold for more general inputs. 

\section{Non-vacuum inputs}
\subsection{Two coherent states}

In this section, we generalize the above results such that both inputs are non-vacuum. 
We consider the two-parameter estimation, i.e. $\phi_s$ and $\phi_d$ are both assumed to be unknown. 
For simplicity, we assume $\theta =0$. 
When the input state is a tensor of two coherent states, $\ket{\alpha}_A\otimes\ket{\beta}_B$, where $\alpha$ and $\beta$ 
are both complex, following the QFIM approach in the previous section, we get 
\begin{equation}
\Delta^2 \phi_s \ge F_{\rm coh}^{-1} ,
\end{equation}
where 
\begin{align}
\label{eqn:QFIM_two_coh}
 F_{\rm coh} &=\frac{{n_{{\rm{in}}}^2{n_\kappa }({n_\kappa } + 2) + 4{n_a}{n_b}{{({n_\kappa } + 1)}^2}}}{{{n_{{\rm{in}}}}}} 
\\ \nonumber
&+ {n_\kappa }({n_\kappa } + 2) + 2{\rm Re} (\alpha \beta) \sinh(4g), 
\end{align}
and $n_{\textrm{in}} = \abs{\alpha}^2+\abs{\beta}^2$. 
If $\abs{\alpha}^2=0$ or $\abs{\beta}^2=0$, we get back \eqnref{eq:trueQFI}. 
Note that \eqnref{eqn:QFIM_two_coh} is always tighter than the one in Ref.~\cite{Gong2016}. 
This discrepancy comes from the fact that Ref.~\cite{Gong2016} treated the problem effectively as a single-parameter estimation (i.e. implicitly assuming that $\phi_d$ is known a priori). 
Also, when $n_\kappa = 0$, and there is no OPA, then the two arms never interact before the measurement step, and we get
\begin{equation}
 F_{\rm coh}^{n_\kappa \to 0} =\frac{4\abs{\alpha}^2\abs{\beta}^2}{\abs{\alpha}^2+\abs{\beta}^2}.
\end{equation}
This result matches with the ``one-mode'' interferometer model described in Ref.~\cite{Marcin2016}. 

For given $n_\kappa$ and $n_{\rm in}$, $F_{\rm coh}$ is maximized when two inputs have the same amplitude and 
conjugate phase, i.e. $\alpha = |\alpha|e^{i \varphi}$ and $\beta = |\beta|e^{-i \varphi}$ with $|\alpha| = |\beta|$.  
Then Eq.~(\ref{eqn:QFIM_two_coh}) reduces to be 
\begin{align}
\label{eqn:QFIM_two_coh_max}
 F^{\rm max}_{\rm coh} & = (n_{\rm in} +1) n_\kappa (n_\kappa + 2) + n_{\rm in} (n_\kappa +1)^2  \\ \nonumber
&+ 4 n_{\rm in} \sqrt{n_\kappa (n_\kappa + 1)} (2n_\kappa +1), 
\end{align}
where note that $\sinh(4g) = 4 \sqrt{n_\kappa (n_\kappa + 1)} (2n_\kappa +1)$. 
We observe that $F^{\rm max}_{\rm coh}$ is basically proportional to $n_\kappa^2$ and $n_{\rm in}$. 
Thus, if we restrict only the total input power to the system $n_{\rm tot} = n_{\rm \kappa} + n_{\rm in}$, 
it suggests one should concentrate all power to the parametric gain to maximize the quantum advantage beyond the shot noise limit. 
If instead we have a restriction on $n_{\rm \kappa}$ due to practical reasons, the coherent state input still can boost the sensitivity by a factor proportional to $n_{\rm in}$.

\subsection{Coherent state and squeezed vacuum}

We further extend the analysis for the input of coherent state and squeezed vacuum: $|{\alpha}\rangle_A \otimes | {\xi}\rangle_B$. 
For simplicity, we assume $\alpha$ is real. 
By using QFIM method, we get the estimation of the phase sum $\phi_{s}$ as in \eqnref{eqn:SU11_QFIM}, is bounded by the QFI: 

\begin{align}
\label{eqn:f_Q^1}
 F_Q^1 &= \sinh ^2(2g) \left[\abs{\alpha}^2 \e^{2 r} +\cosh ^2(r)\right]\\ \nonumber
 &+ \cosh ^2(2g) \frac{8 \abs{\alpha}^2 \sinh ^2(2r)}{4 \abs{\alpha}^2+2 \sinh ^2(2r)},
\end{align}
where $r$ is the squeezing strength of the squeezed vacuum $\ket{\xi}$. When any or both of the input state is vacuum, i.e, $r=0$ or $\abs{\alpha}^2=0$, we get back \eqnref{eq:trueQFI}. In addition, when $g=0$, this means that the two modes don't interact before the measurement step, and when either $r=0$ or $\abs{\alpha}^2=0$, we get $F=0$, which is consistent with our intuition. 

It is worthwhile to compare the above result with that of Li \textit{et al.}'s. The QFI in Ref. \cite{Li2016} is given by: 
\begin{align}
 F_Q^2 &=\sinh^2(2g) \left[\abs{\alpha}^2 \e^{2 r}+\cosh^2(r)\right]\\ \nonumber
 &+\cosh^2(2g) \left[\abs{\alpha}^2+\frac{1}{2} \sinh^2(2r)\right].
\end{align}

The difference between $F_Q^1$ and $F_Q^2$ is given by: 
\begin{equation}
 F_Q^1-F_Q^2=- \cosh ^2{2 g}\frac{ \left[-4 \abs{\alpha}^2+\cosh(4 r)-1\right]^2}{4 \left[4 \abs{\alpha}^2+\cosh(4r)-1\right]},
\end{equation}
We can see the difference is always negative, which suggests that our phase sensitivity,given by $F_Q^1$, provides a tighter QCRB.

It is also interesting to compare the bound in Eq.~(\ref{eqn:f_Q^1}) with the classical Fisher information (CFI) of parity detection, which is the known best strategy in ideal scenario in this setup \cite{Li2016}. The CFI of parity detection is given by 
\begin{equation}
 F_{\textrm{cl}}=\sinh^2(2g)\left[\abs{\alpha}^2 \e^ {2 r}+\cosh^2(r)\right].
\end{equation}
Obviously, $F_Q^1$ is larger than $F_{\textrm{cl}}$, which suggests that the parity measurement is indeed not the optimal measurement in this case. Note that for multi-parameter estimation problems, there is no guarantee that there exists a set of POVM which saturates the QCRB \cite{Helstrom1976,Matsumoto2002,Sammy2016,luca2017,yang2018optimal}.

\section{Conclusion}
In this paper, we revisited the quantum Fisher information approach to establish a fundamental precision estimation bounds for SU(1,1) interferossmeters. Firstly, when one of the input state is restricted to be a vacuum state, we showed that by using either the phase-averaging method or the quantum Fisher information matrix method, different phase configurations of the SU(1,1) interferometer result in the same QFI. In this case, the QFI is linearly proportional to the average photon number of the second input state, and quadratically proportional to the average photon number generated by the OPA. This suggests that when fixing the squeezing strength of the OPA, to achieve higher sensitivity, one simply needs to inject a state with higher average photon number. Secondly, we compared the results of the phase-averaging method and the quantum Fisher information matrix method, and then we argued that for a SU(1,1) interferometer, phase averaging or quantum Fisher information matrix method is generally required, and they are essentially equivalent. Finally, we used the quantum Fisher information matrix method to calculate the precision limit for other common input states, such as two coherent state inputs or coherent state with squeezed vacuum inputs. 

\section*{ACKNOWLEDGEMENT}
CY would like to acknowledge support from Economic Development Assistantship from Louisiana State University System Board of Regents. SA and JPD would like to acknowledge support from Army Research Office and National Science Foundation. XPM acknowledges financial support from National Natural Science Foundation of China. 

\bibliographystyle{apsrev4-1}
\bibliography{Phase_distribution}

\end{document}